\documentclass[aps,prb,twocolumn,showpacs,preprintnumbers,amsmath,amssymb]{revtex4}
\usepackage{graphicx}
\usepackage{dcolumn}
\usepackage{bm}

\begin{document}
\title{Equation of motion for dislocations with inertial effects}

\author{L.\ Pillon}
\author{ C.\ Denoual}
\email{christophe.denoual@cea.fr}
\author{ Y.-P. Pellegrini}
\email{yves-patrick.pellegrini@cea.fr}
\affiliation{D\'epartement de Physique Th\'eorique et Appliqu\'ee, \\
        Commissariat \`a l'\'Energie Atomique, BP12\\
        F-91680 Bruy\`eres-le-Ch\^atel, France.}

\date{\today}

\begin{abstract}
An approximate equation of motion is proposed for screw and edge dislocations,
which accounts for retardation and for relativistic effects in the subsonic
range. Good quantitative agreement is found, in accelerated or in decelerated
regimes, with numerical results of a more fundamental nature.
\end{abstract}


\maketitle

\section{Introduction}
Dislocation  behavior in solids under dynamic conditions (e.g.\ shock loading
\cite{HORN62,CLIF81,TANG03}) has recently attracted renewed attention,
\cite{TANG03,HIRT98,GUMB99,MARI04,DENO04,PILL06,DENO07} partly due to new
insights provided by molecular dynamics studies. \cite{GUMB99,TANG03,MARI04}
Whereas theoretical investigations mainly focused on the stationary velocities
that regular or twinning dislocations can attain as a function of the applied
stress (possibly intersonic or even supersonic with respect to the longitudinal
wave speed $c_\text{L}$),\cite{WEER69,GUMB99,ROSA01} one other major concern is
to establish an equation of motion \cite{ESHE53,CLIF81,ROSA01,HIRT98} (EoM)
suitable to instationary dislocation motions towards or from such high
velocities, and which is computationally cheap. This would be an important step
towards extending dislocation dynamics (DD) simulations
\cite{DEVI97,BULA06,WANG06} to the domain of  high strain rates, in order to
better understand hardening processes in such conditions.

The key to instationary motion of dislocations lies in the inertia arising from
changes in their long-ranged displacement field, which accompany the motion.
These retarded rearrangements take place at finite speed, through wave emission
and propagation from the dislocation. As a consequence, dislocations possess an
effective inertial mass,\cite{ESHE53} which has bearings on the process of
overcoming dynamically obstacles such as dipoles, etc. \cite{PILL06,WANG06,BITZ05}
Inertial effects are non-local in time, and are related to effective
``viscous'' losses. For small velocities where the EoM is linear,\cite{ESHE53}
this relation takes the form of the Kramers-Kr\"onig relations between the
reactive and dissipative parts of the causal mass
kernel.\cite{NABA51,ESHE62,ALSH71} One major ingredient of the EoM should thus
be the effective visco-inertial force exerted on the dislocation by its own
wave emission.\cite{ESHE53,CLIF81} An EoM results from balancing it by the
applied stress, and by drags of various origins.\cite{ALSH86} EoMs with
effective masses, but which ignore retardation (e.g., Ref.\
\onlinecite{HIRT98}), cannot truly capture visco-inertial effects. Previous
works on these questions having mainly been confined to the linear regime,
their influence in the relativistic domain remains largely unexplored in spite
of analytical progresses, partly due to the complexity of the formalism
(especially for edge dislocations).

Hereafter, Eshelby's EoM for screws with a rigid core,\cite{ESHE53} valid at
small velocities, is first re-examined, and cast under a simple form which
suggests a straightforward regularization procedure for finite core effects.
This allows us to appeal to previous results for point dislocations valid at
high velocities.\cite{CLIF81} We then build in an heuristic way an EoM for
accelerated or decelerated screw and edge dislocations in the drag-dominated
subsonic regime, that consistently accounts for saturation effects at
velocities comparable to the sound speed. Results from the equation are
compared to quasi-exact calculations from a numerical method of the phase-field
type. Having in mind applications to DD simulations, the scope of the study is
limited to continuum theory, so that dispersion effects due to the atomic
lattice,\cite{ISHI73} or to the influence of the Peierls
potential,\cite{ALSH71} are not explicitly considered.

\section{Eshelby's force}
Within the Peierls-Nabarro model in isotropic elasticity,\cite{PEIE40,NABA47}
and with the usual $\arctan$ ansatz for the relative displacement $u(x,t)$ of
the atoms on both sides of the glide plane, Eshelby computed the visco-inertial
force $F$ experienced by a screw dislocation of Burgers vector $b$, centered on
position $x(t)$ at time $t$, moving with a velocity $v=\dot{x}$ small compared
to the shear wave speed $c_\text{S}$:\cite{ESHE53}
\begin{eqnarray}
\label{eq:eshforce}&&F(t)=m_0\int_{-\infty}^t{\rm d}\!\tau
\frac{\dot{v}(\tau)}{\left[(t-\tau)^2+t_{\rm S}^2\right]^{1/2}}\\
&&{}+m_0\int_{-\infty}^t \hspace{-1em}{\rm d}\!\tau \frac{t_{\rm
S}^2}{\left[(t-\tau)^2+t_{\rm S}^2\right]^{3/2}} \frac{\rm d}{{\rm
d}\tau}\left(\frac{x(t)-x(\tau)}{t-\tau}\right)\nonumber.
\end{eqnarray}
The dislocation is assumed to have a \emph{rigid} core of half-width $\zeta_0$.
Then $t_{\rm S}=2\zeta_0/c_\text{S}$ is the time of shear wave propagation over
the core width. The mass per unit dislocation length $m_0=\mu b^2/(4\pi
c_\text{S}^2)$ depends on the shear modulus $\mu$. In Ref.\ \onlinecite{ESHE53}
(and in Ref.\ \onlinecite{ALSH71} as well), an incorrect factor $1/2$ is
present in front of the second integral, and has been removed here. This factor
is of no important physical consequence, save for different values of the
linear response kernels; see below.

That (\ref{eq:eshforce}) is correct can be verified as follows. Starting from
Eshelby's expression of the force as a double integral in Eq.~(26) of Ref.\
\onlinecite{ESHE53}, and expanding it to linear order in the velocity $v(\tau)$
or in $x(t)-x(\tau)$, the following expression is easily obtained:
\begin{eqnarray} \label{eq:flin1}
F(t) &=& 2 m_0\int_{-\infty}^t {\rm
d}\tau\,\left\{\frac{\dot{v}(\tau)}{\sqrt{\Delta
t^2+t_{\rm S}^2}}\right.\nonumber\\
&&\left. {}-\frac{2 t_{\rm S}^2-\Delta t^2}{(\Delta t^2+t_{\rm
S}^2)^{5/2}}[x(t)-x(\tau)]\right\},
\end{eqnarray}
where $\Delta t=t-\tau$. Using integrations by parts over $\tau$, each of
Eq.~(\ref{eq:eshforce}) and (\ref{eq:flin1}) can be put under the following
irreducible form:
\begin{equation}
\label{eq:canonical} \frac{F(t)}{m_0}=2\frac{v(t)}{t_{\rm
S}}-2\frac{x(t)}{t_{\rm S}^2}+2\int_{-\infty}^t {\rm d}\tau\,\frac{
x(\tau)}{\left[(t-\tau)^2+t_{\rm S}^2\right]^{3/2}},
\end{equation}
which shows them to coincide.

By the same token, we check that
(\ref{eq:eshforce}) can be further simplified as:
\begin{equation}
\label{eq:flin2} F(t)=\int_{-\infty}^t{\rm d}\!\tau \frac{2
m_0}{\left[(t-\tau)^2+t_{\rm S}^2\right]^{1/2}}\frac{\rm d}{{\rm
d}\tau}\left(\frac{x(t)-x(\tau)}{t-\tau}\right).
\end{equation}
By Fourier transforming $F(t)$ [under the form (\ref{eq:canonical})] and by
writing
$$F(\omega)\equiv [-\omega^2 m(\omega)-i\omega \eta(\omega)]\,x(\omega),$$
we identify effective mass $m(\omega)$ and viscosity $\eta(\omega)$ kernels.
\cite{NABA51} Their expression in closed form involves the modified Bessel and
Struve functions $K_1$, $I_1$ and $\mathcal{L}_{-1}$:
\begin{subequations}
\label{eq:masskercf}
\begin{eqnarray}
\label{meshcf} \frac{m(\omega)}{m_0}&=&2\frac{1-t_{\rm S}|\omega|K_1(t_{\rm S}|\omega|)}{(t_{\rm S}|\omega|)^2}\\
\label{veshcf}\frac{\eta(\omega)}{m_0}&=&\frac{2}{t_{\rm
S}}\left\{1+\frac{\pi}{2}\bigl[I_1(t_{\rm S}|\omega|)-\mathcal{L}_{-1}(t_{\rm
S}|\omega|)\bigr]\right\}
\end{eqnarray}
\end{subequations}
To leading orders in the pulsation $\omega$,
\begin{subequations}
\label{eq:massker}
\begin{eqnarray}
\label{mesh} m(\omega)/m_0&=&\left(\frac{1}{2}+\ln\frac{2e^{-\gamma}}{t_{\rm
S}|\omega| }
\right)+O\left((t_{\rm S}\omega)^2\ln t_{\rm S}\omega\right)\\
\label{vesh} \eta(\omega)/m_0&=&\frac{\pi}{2}|\omega|+O(t_{\rm S}|\omega|^2)
\end{eqnarray}
\end{subequations}
where $\gamma$ is Euler's constant. Moreover, we observe that
\begin{equation}
\label{eq:viscoinfty} \eta(|\omega|\to\infty)/m_0=2/t_{\rm S}.
\end{equation}
Result (\ref{eq:massker}) coincides to leading order with Eshelby's,
\cite{ESHE53} as $\omega\to 0$. The mass increase with wavelength as $\omega\to
0$ implies very different behaviors for, e.g., quasi-static and shock loading
modes, since the latter involves a wider frequency range. We note that
$\eta(\omega)\to 0$ as $\omega\to 0$, since losses should be absent from the
model in the stationary subsonic regime. \cite{ESHE53} The non-analytical
behavior of the kernels at $\omega=0$ (due to $|\omega|$), and its associated
non-locality in time has been emphasized in Ref.\ \onlinecite{ALSH71}.

The finite ``instantaneous'' viscosity (\ref{eq:viscoinfty}) stems from the
first term in the R.H.S.\ of (\ref{eq:canonical}), and is responsible for a
velocity jump $\Delta v$ undergone by the dislocation when subjected to a jump
$\Delta F$ in the applied force.\cite{ESHE53,CLIF81} From (\ref{eq:viscoinfty})
we deduce:
\begin{equation}
\label{eq:vjump} \Delta v=\frac{\Delta F}{\eta(\infty)}=\frac{t_{\rm S}\Delta
F}{2 m_0}=4\pi\frac{\zeta_0 c_\text{S}\Delta F}{\mu b^2}.
\end{equation}
The velocity jump (\ref{eq:vjump}) increases with core width. It was first
predicted by Eshelby from his equation,\cite{ESHE53} and can be understood as
follows for a screw dislocation along the $z$ axis: the force jump $\Delta F$
is due to a shear stress jump $\Delta\sigma_{yz}=\Delta F/b$ attaining
simultaneously all the points of the whole glide plane (e.g., as the result of
shear loading applied on faces of the system containing the plane, parallel to
the latter). Neglecting material inertia of the atoms on both sides of the
dislocation plane, the medium undergoes an elastic strain jump $\Delta
\sigma_{yz}/\mu=\Delta \varepsilon_{yz}\sim\Delta v_m/c_\text{S}$, determined
by a material velocity jump $\Delta v_m$. The latter is equilibrated through
outward emission of a shear wave with velocity $c_\text{S}$. On the other hand,
the slope of the displacement function near the core is $(\partial u/\partial
x)\sim b/(2\zeta_0)$, so that $\Delta v_m$ is related to the dislocation
velocity jump $\Delta v$ by $\Delta v_m\sim \Delta v\, b/(2\zeta_0)$. Combining
these relationships yields (\ref{eq:vjump}), up to a numerical constant factor.
The same argument applies to other types of dislocations. In case of several
relaxation waves (e.g., longitudinal and shear waves for an edge dislocation),
that of lowest celerity controls the amplitude of the velocity jump. It should
be borne in mind, however, that accounting for material inertia from the atoms
on both sides of the glide plane results in an instantaneous inertial force of
order $F_i=2m_0 \ddot{x}$ to be added to (\ref{eq:eshforce}).\cite{ESHE53} By
balancing the forces, it is seen that this force should spread the velocity
jump over a short rise time
\begin{equation}
\label{eq:risetime} \Delta t \sim t_{\rm S}.
\end{equation}

\section{Equation of motion}
No expression analogous to (\ref{eq:eshforce}) is available for edge
dislocations. However, Clifton and Markenscoff computed the force acting on a
\emph{point} screw or edge dislocation moving with any subsonic velocity in an
isotropic medium, that jumps instantaneously at instant $t=\tau$ from rest to a
constant velocity $v$.\cite{CLIF81} A generalization to anisotropic media is
available.\cite{WU02} To maintain its velocity constant, this dislocation must
be subjected, at time $t>\tau$, to the time-decaying force
\begin{equation}
\label{eq:fcm} F^{\rm CM}(t-\tau,v)=\frac{g\bigl(v\bigr)}{t-\tau},
\end{equation}
where the function $g$ depends on its character and on anisotropy.
\cite{ESHE53,CLIF81} We now construct heuristically a force for accelerated
motion by interpreting such a motion as a succession of infinitesimal velocity
jumps. Assuming that, for instationary motion, $v$ in (\ref{eq:fcm}) can be
interpreted as $v(\tau)$, the elementary force that would arise from the
elementary jump $\delta v(\tau)$ at $t=\tau$ is: $ \delta F=[\partial F^{\rm
CM}\bigl(t-\tau,v(\tau)\bigr)/\partial v(\tau)]\delta v(\tau)$ $=$
$g'\bigl(v(\tau)\bigr)\delta v(\tau)/(t-\tau)$. Then, the total force
experienced by the dislocation results from integrating such elementary forces
over past history:
\begin{equation}
\label{eq:newforce0} F(t)=\int_{-\infty}^t {\rm d}\!\tau\, \frac
{g'\bigl(v(\tau)\bigr)}{t-\tau}\dot{v}(\tau).
\end{equation}
Comparing (\ref{eq:newforce0}) to (\ref{eq:flin2}) shows, firstly, that the
relevant ``accelerations'' at linear order are different. However, we remark
that $2({\rm d}/{\rm d}t)\{[x(t)-x(\tau)]/(t-\tau)\}\to \dot{v}(\tau)$ as
$t\to\tau$, and moreover that for a screw dislocation, $g'(v\simeq 0)=m_0$.
\cite{CLIF81} Hence, since we interpret $v$ in (\ref{eq:fcm}) as $v(\tau)$, the
numerator of the integrand in (\ref{eq:newforce0}) is correct at least for
small velocities and for small times $t\to\tau$. Its relevance for large
velocities is demonstrated below through comparisons to full-field
calculations. Next, integral (\ref{eq:newforce0}) is singular at $\tau=t$, due
the point-dislocation hypothesis at the root of (\ref{eq:fcm}). However, using
(\ref{eq:flin2}) as a physical motivation, we propose a regularization
consisting in replacing the kernel $1/t$ in (\ref{eq:newforce0}) by
$1/[t^2+t_0^2]^{1/2}$ where $t_0$, the counterpart of $t_{\rm S}$ in
(\ref{eq:eshforce}), is some time characteristic of sound propagation over a
core diameter. In Sec.\ \ref{sec:appli}, $t_0$ is chosen alternatively
proportional to $t_{\rm S}=2\zeta_0/c_\text{S}$ and to $t_{\rm
L}=2\zeta_0/c_\text{L}$ in the case of edge dislocations for illustrative
purposes, whereas $t_0$ is proportional to $t_{\rm S}$ for screws. The
proportionality factor, 1/2 in all cases, is justified below. From a physical
point of view, inertia is controlled by the slowest wave so that better results
are expected using $c_\text{S}$ for all types of dislocations. Given
Eshelby's rigid-core hypothesis in (\ref{eq:flin2}), and the approximations
made, it would be pointless to refine this treatment. Another kind of
regularization is used in Ref.\ \onlinecite{HIRT82} (p.\ 195), which consists
in replacing the upper bound $t$ of integral (\ref{eq:newforce0}) by $t-t_0$
(in Ref.\ \onlinecite{HIRT82}, the integrand assumes that $v\simeq 0$).

With the above regularization the force eventually reads:
\begin{equation}
\label{eq:newforce1} F_{\text{reg}}(t)=\int_{-\infty}^t {\rm d}\!\tau\, \frac
{g'\bigl(v(\tau)\bigr)}{[(t-\tau)^2+t_0^2]^{1/2}}\dot{v}(\tau).
\end{equation}
Its Fourier transform for small velocities where $g'(v)\simeq g'(0)$ yields, in
terms of modified Bessel and Struve functions of order 0,
\begin{subequations}
\label{eq:masskercfapprox}
\begin{eqnarray}
\label{meshcfapprox} m(\omega)/g'(0)&=&K_0(t_0|\omega|)\\
&=&\ln\frac{2e^{-\gamma}}{t_{\rm S}|\omega|}
+O\left((t_{\rm S}\omega)^2\ln t_{\rm S}\omega\right),\nonumber\\
 \label{veshcfapprox}\eta(\omega)/g'(0)&=&\frac{\pi}{2}|\omega|
\bigl[I_0(t_0|\omega|)-\mathcal{L}_{0}(t_0|\omega|)\bigr]\nonumber\\
&=&\frac{\pi}{2}|\omega|+O(t_{\rm S}|\omega|^2)\\
 \label{veshcfapproxinf}\eta(|\omega|\to\infty)/g'(0)&=&1/t_0.
\end{eqnarray}
\end{subequations}
The approximation therefore preserves the logarithmic character of the mass,
and the viscosity, to leading order. The mass is slightly decreased, the
constant $m_0/2$ in (\ref{mesh}) being absent. This difference is insignificant
given the approximations made. In the limit of small velocity for a screw
dislocation, our approximation amounts to retaining in (\ref{eq:eshforce}) the
first integral only. In order to recover a correct velocity jump for screws, we
must take $t_0\simeq t_{\rm S}/2$ since the instantaneous viscosity
(\ref{veshcfapproxinf}) is different from (\ref{eq:viscoinfty}). This
``calibration'' is used in the
 next section for screws and (somewhat arbitrarily) for edges as well.

In the stationary limit, the visco-inertial force (\ref{eq:newforce1})
vanishes. For $v\ll c_{\rm S}$, the asymptotic velocity should be determined by
a viscous drag force, mainly of phonon origin,\cite{ALSH86} $F_{\rm
drag\,0}=\eta_0 v$, where $\eta_0$ is the viscosity. This force is modified (in
the context of the Peierls-Nabarro model) by the relativistic contraction of
the core, into $F_{\rm drag}(v)=\eta(v) v$. For subsonic velocities,
$\eta(v)=\eta_0 D(0)/D(v)$, where:\cite{ROSA01}
 \begin{equation}
 \label{eq:contract}
 D(v)=\left[A^2(v)+\alpha^2 (v/c_{\rm S})^2 \right]^{1/2},
\end{equation}
with $\alpha=\eta_0\zeta_0/(2 m_0 c_{\rm S})$, is an effective
viscosity-dependent core contraction factor, such that the core length in the
laboratory frame reads: $\zeta(v)=\zeta_0 D(v)/D(0)$. The purely relativistic
contraction factor $A(v)$ is, with $\gamma_{\rm L,S}=\left(1-v^2/c_{\rm
L,S}^2\right)^{1/2}$:\cite{WEER61,ESHE49,ROSA01}
\begin{equation}
A(v)= \left\{
\begin{array}{c}
\frac{1}{2}(c_{\rm S}/v)^2\left(4\gamma_{\rm L}
-\gamma_{\rm S}^{-1}-2\gamma_{\rm S}-\gamma_{\rm S}^{3}\right)\mbox{ for edges},\\
\vphantom{\Bigl(}\frac{1}{2}\gamma_{\rm S}\hspace{12em}\mbox{ for screws}.
\end{array}
\right.\nonumber
\end{equation}
With this drag, and introducing the applied stress $\sigma_a$, the EoM finally
reads:
\begin{equation}
\label{eq:pdpeq} \frac{\mu b^2}{2\pi}\int_{-\infty}^t{\rm d}\!\tau\,
\frac{\widetilde{g}'\bigl(v(\tau)\bigr)\dot{v}(\tau)}{[(t-\tau)^2+t_0^2]^{1/2}}+F_{\rm
drag}(v(t))=b\sigma_a,
\end{equation}
where $g(v)\equiv 2 m_0\, \widetilde{g}(v)$, and where:\cite{CLIF81}
\begin{eqnarray*}
\widetilde{g}(v)&=&(\gamma_{\rm S}^{-1}-1)/v,\hspace{3.5em}\text{for screw dislocations},\\
&=&(8\gamma_{\rm L}+4\gamma_{\rm L}^{-1}-7\gamma_{\rm S}-6\gamma_{\rm S}^{-1}+\gamma_{\rm S}^{-3})c_{\rm S}^2/v^3\nonumber\\
&&-2[1-(c_{\rm S}^2/c_{\rm L}^2)^2]/v,\quad\text{for edge dislocations}.
\end{eqnarray*}
This is our main result. By construction, it reproduces the asymptotic
velocities of Ref.\ \onlinecite{ROSA01}.

We checked numerically that the replacement of $\zeta_0$ by $\zeta(v)$ in $t_0$
does not change by more than a few percent the overall results described in the
following section. Since this change in $t_0$ would bring in nothing useful, we
choose to use $\zeta_0$ in $t_0$ in the following section.
\begin{figure}
\includegraphics[width=8cm]{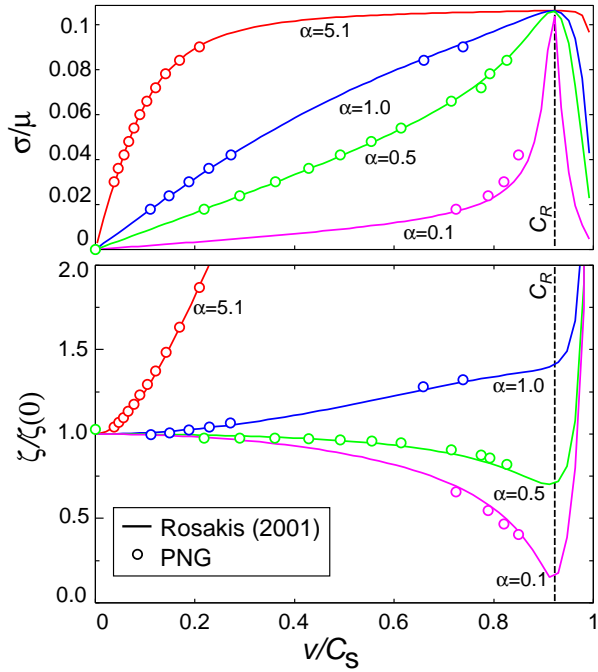}
\caption{\label{fig:rosakis} Upper: Relationship between dimensionless applied
stress $\sigma/\mu$ and asymptotic velocity $v/c_\text{S}$ provided by the PNG
code, \cite{DENO07} for an accelerated edge dislocation in the stationary
regime of an accelerated dislocation (dots), compared to that predicted by
Rosakis's Model I\cite{ROSA01} (lines) for different viscosity parameters
$\alpha$, for a screw dislocation in the subsonic regime. Lower: Normalized
velocity-dependent core width $\zeta(v)/\zeta(0)$ measured under same
conditions. $c_\text{R}$ is the Rayleigh velocity.}
\end{figure}

\section{Applications}
\label{sec:appli} Setting $v(t)=\sum_i \Delta v_i \theta(t-t_i)$,
Eq.~(\ref{eq:pdpeq}) is solved numerically for edge and screw dislocations, in
an implicit way with a time step $\Delta t=t_{i+1}-t_i$ small enough. Results
are compared with numerical points obtained with the
\emph{Peierls-Nabarro-Galerkin} (PNG) approach \cite{DENO04,DENO07} used here
as a benchmark. This method is less noisy than molecular dynamics, allows for
full-field dynamic calculations of the displacement and stress fields in the
whole system, accounting for wave propagation effects, and allows for better
flexibility. We can thus, e.g., control boundary conditions by applying
analytically computed forces, so as to prevent image dislocations from
perturbing the simulation window.

Firstly, to check the accuracy of the benchmark, asymptotic velocities of screw
and edge dislocations were compared to the stationary predictions of Rosakis'
Model 1.\cite{ROSA01} In the PNG method, the permanent lattice displacement
field (which is part of the full atomic displacement, $u$) is relaxed by means
of a Landau-Ginzburg equation, with viscosity parameter $\eta_\text{PNG}$. An
exact correspondence holds between this viscosity and Rosakis's viscosity
parameter $\alpha$, namely $\eta_\text{PNG}=\alpha\mu/c_\text{S}$, as can be
shown by specializing to one dimension the general field equations of Ref.\
\onlinecite{DENO04}. A $\gamma$-potential $\gamma(u)=(1/2)\gamma_0\sin^2(\pi
u/b)$, with
 $\gamma_0=(2/3)C_{44}b/\pi^2$, is used. The material is an
elastically cubic material, with elastic moduli taken such that
$C_{44}=C_{12}=C_{11}/3$ to insure isotropy. Due to the elastic correction made
to the $\gamma$-surface potential in order to remove its quadratic elastic
part,\cite{DENO04,DENO07} the core at rest is a bit larger in the PNG results
than in the Peierls-Nabarro solution. The time dependent core width
$\zeta\bigl(v(t)\bigr)$ is measured from the numerical simulations by using
$b^2/(2\pi\zeta)\equiv \int {\rm d}x\, [u'(x)]^2$ (the value corresponding to a
core of the arctan type). Two-dimensional calculations are carried out using a
simulation box of size $300\times 30$ $b^2$, with a unique horizontal glide
plane along $Ox$. Eight nodes per Burgers vector are used in both directions.
Forces are applied on the top and bottom sides so as to induce shear on the
unique glide plane. Free boundary conditions are used on sides normal to the
$Ox$ axis. Measurements are done near the center of the box, where the mirror
attracting forces these sides generate on the dislocation, are negligible. The
box is wide enough so that the dislocation accelerates and reaches its terminal
velocity. Comparisons between PNG results and Rosakis' model are displayed in
Fig.\ \ref{fig:rosakis} for different viscosities $\alpha$, in the case of an
edge dislocation. The core scaling factor $D(v)$ and the asymptotic velocity
$v/c_\text{S}$, are directly measured from simulations under different applied
stresses $\sigma$, and compared to theory. \cite{ROSA01} The PNG asymptotic
velocities were found to be $5\%$ systematically lower than the theoretical
results. This correction is accounted for in the figure. The overall agreement
is excellent. It is emphasized that core contraction effects in the viscous
drag [Eq.\ (\ref{eq:contract})] are required in order to obtain a good match.
\begin{figure}
\includegraphics[width=8.5cm]{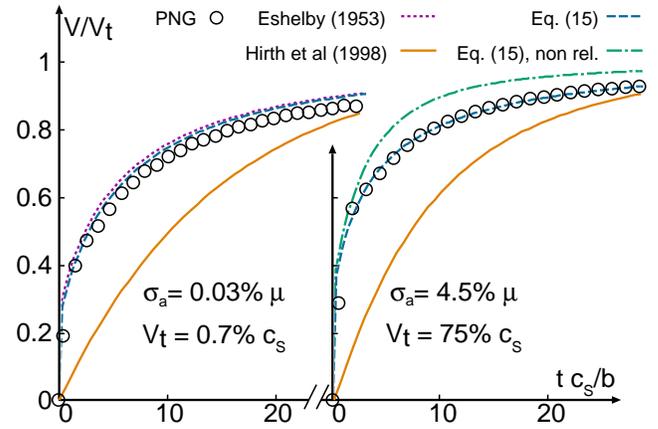}
\caption{\label{fig:step_vis}Velocities vs.\ time for accelerated screw
dislocations: white dots, PNG code; solid, Ref.\ \onlinecite{HIRT98}; dots (in
left curve only), Eq.~(\ref{eq:eshforce}); dash-dots, linear approximation to
(\ref{eq:pdpeq}); dashes, fully relativistic equation (\ref{eq:pdpeq}).}
\end{figure}
\begin{figure}
\includegraphics[width=8.5cm]{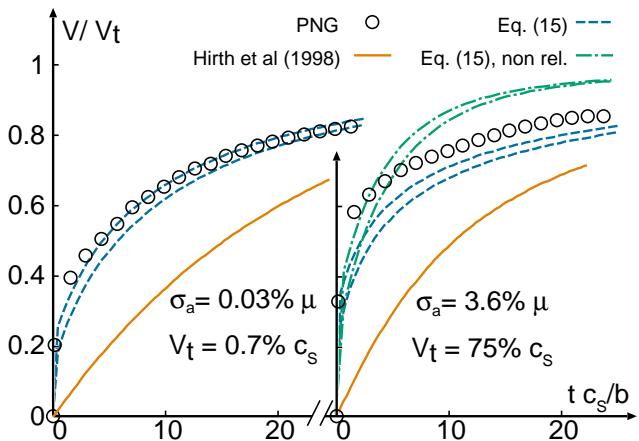}
\caption{\label{fig:step_coin}Velocities vs.\ time for accelerated edge
dislocations: white dots, PNG code; solid, Ref.\ \onlinecite{HIRT98};
dash-dots, linear approximation to (\ref{eq:pdpeq}); dashes, fully relativistic
equation (\ref{eq:pdpeq}). Curves obtained from (\ref{eq:pdpeq}) are
duplicated, using either $t_0$ computed with $c_\text{S}$ (upper), or with
$c_\text{L}$ (lower), see text.}
\end{figure}

Next, comparisons in the accelerated regime are made with Eq.~(\ref{eq:pdpeq})
and with other models. Fig.\ \ref{fig:step_vis} displays, as a function of
time, the velocity of a screw dislocation accelerated from rest by a constant
shear stress $\sigma_a$ applied at $t=0$. Low and high shear stresses are
examined. These stresses lead to terminal asymptotic velocities
$v_t=v(t=\infty)=0.007\, c_{S}$ and $0.75\, c_{S}$, computed from
(\ref{eq:pdpeq}). The results displayed are obtained: (i) with the PNG approach
(white dots); (ii) with Eq.~(\ref{eq:pdpeq}) using fully ``relativistic''
expressions of $\widetilde{g}(v)$ and $D(v)$ (dashes); (iii) with linear
small-velocity approximations of $\widetilde{g}(v)$, but with the full
expression of $D(v)$, in order to emphasize the importance of relativistic
effects in the retarded force (dash-dots, for the case $v_\text{T}=0.75
c_\text{S}$); (iv) with a previous EoM,\cite{HIRT98} using a typical cut-off
radius $R=500$ nm in the logarithmic core term (solid) corresponding to a
typical dislocation density of $10^{12}$/m${}^2$. The result arising from using
(\ref{eq:eshforce}) in the EoM is also displayed for the lowest speed (dots,
left figure only).

\begin{figure}
\includegraphics[width=8.5cm]{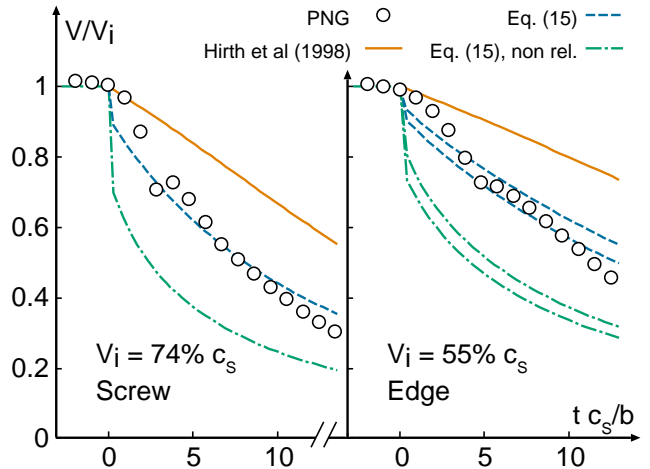}
\caption{\label{fig:deceler}Velocities vs.\ time for a decelerated screw
dislocation. Comparison between the PNG method (white dots)\cite{DENO04} and
equation (\ref{eq:pdpeq}) with (dashes) or without (dot-dash) fully
relativistic expressions.}
\end{figure}

Figure \ref{fig:step_coin} presents similar curves for an edge dislocation. For
the latter, $t_0$ is taken either as $t_{\rm S}/2$ or as $t_{\rm L}/2$, $t_{\rm
L}=2\zeta_0/c_\text{L}$, thus providing two limiting curves. The curves with
$t_{\rm S}$ provide the best matches, consistently with the above observation
that the wave of lowest velocity $c_{\rm S}<c_{\rm L}$ should provide the main
contribution to inertia. At low and high speeds, good agreement is obtained
between PNG points and Eq.~(\ref{eq:pdpeq}), provided that fully
``relativistic'' expressions are used for $g(v)$ (especially for edge
dislocations); otherwise, inertia is strongly underestimated. In all the
curves, the relativistic expression of the non-linear viscous terms was used.
Moreover, variations of the core width with velocity,\cite{ROSA01} implicitly
present in PNG calculations, and ignored in the expression of $t_0$ used in the
visco-inertial term of (\ref{eq:pdpeq}), are not crucial to accelerated or to
decelerated motion (see Fig.\ \ref{fig:deceler}); still, the core width shrinks
by 20\% during the acceleration towards $v_t=0.75\, c_{L}$. On the other hand,
retardation effects in the effective mass are crucial: curves with non-local
inertial forces are markedly different from the solid ones using the masses of
Ref.\ \onlinecite{HIRT98}, computed at constant velocity. The version of the
PNG code used here does not include the above-mentioned effects of material
inertia in the glide plane, so that the full-field velocity curves indeed
display what resembles a velocity jump, like the EoM.  Owing to
(\ref{eq:risetime}), this lack of accuracy solely concerns the time interval
between the time origin and the first data point: hence we can consider that
the velocity jump is a genuine effect, and not an artefact, at least from the
point of view of full-field calculations in continuum mechanics. However, we
should add that, to our knowledge, this effect has not been reported so far in
molecular dynamics simulations.
Figure \ref{fig:deceler} displays the velocity of screw and edge dislocations
decelerated from the initial velocity $v_i$. Comparisons between EoMs and PNG
calculations are then harder to make than in the accelerated case. Indeed, the
non-relativistic (resp.\ relativistic) theoretical curves from Eq.\
(\ref{eq:pdpeq}) (dashed-dot) [resp.\ (dashed)] are obtained by assuming that
an applied stress abruptly vanishes at $t=0$. This induces a negative velocity
jump in the curves. This jump is larger if non-relativistic expressions are
used, which demonstrates in passing the higher inertia (i.e.\ ``mass'')
provided by relativistic expressions. The same loading was tried in the PNG
calculations as well, but led to non-exploitable results due to multiple
wave-propagation and reflection phenomena. Therefore, PNG curves for
decelerated motion were obtained instead using a somewhat artificial loading:
the medium was split in a zone of constant stress, separated from a zone of
zero stress by an immobile and sharp boundary. The dislocation is then made to
accelerate in the zone of constant stress. Due to the finite core width, the
boundary is crossed in a finite time $\simeq \zeta / v_i$, which explains the
smoothed decay of the velocity in the PNG data points. This type of loading
cannot be realistically implemented within the framework of Eq.\
(\ref{eq:pdpeq}) because the dislocation core is not spatially resolved. Hence,
though the curves strongly suggest that relativistic effects are as important
in deceleration as in acceleration, and that (\ref{eq:pdpeq}) reproduces well
the PNG points, the comparison between the latter and theoretical curves should
be taken here with a grain of salt. On the other hand, the EoM of Ref.\
\onlinecite{HIRT98} (solid) is once again clearly imprecise. As a final remark,
we expect our neglecting of retardation effects in the nonlinear viscous term
of (\ref{eq:pdpeq}) to induce an underestimation of damping effects. This may
explain why the PNG curves decay faster than that from Eq.\ (\ref{eq:pdpeq}).

\section{Concluding remarks}
An empirical relativistic equation of motion for screw and edge dislocations,
accounting for retardation effects in inertia, Eq.~(\ref{eq:pdpeq}), has been
proposed. We compared it, together with another available approximate EoM, to a
quasi-exact numerical solution of a dynamical extension of the Peierls-Nabarro
model, provided by the \emph{Peierls-Nabarro Galerkin} code.\cite{DENO07} The
latter was beforehand shown to reproduce very well the asymptotic velocities of
Rosakis's model 1\cite{ROSA01} in the subsonic regime. The best matches with
full-field results were found with our EoM, both for accelerated and for
decelerated motion, thus illustrating quantitatively the importance of
retardation and of relativistic effects in the dynamic motion of dislocations.
To these effects, our EoM provides for the first time a satisfactory
approximation for high velocities in the subsonic range. Our comparisons rule
out the use of masses computed at constant velocity. One of the restrictions
put forward by Eshelby to his EoM was its limitation to weakly accelerated
motion, mainly due to the rigid core assumption.\cite{ESHE53} Ours makes no
attempt to explicitly overcome this simplification. However, comparisons with
full-field calculations, where the core structure is not imposed from the
start, but emerges as the result of solving the evolution equation for the
displacement field, shows that this rigid-core assumption is acceptable on a
quantitative basis as far as inertia is concerned, at least for velocities
high, but not too close to $c_{\rm S}$.

\appendix

\begin{acknowledgments}
The authors thank B.~Devincre for stimulating discussions, and F.\ Bellencontre
for his help during preliminary calculations with the PNG code.
\end{acknowledgments}



\end{document}